\documentstyle[aps,psfig]{revtex}

\newcommand{\bqn}{\begin{eqnarray}}
\newcommand{\eqn}{\end{eqnarray}}
\newcommand{\beq}{\begin{equation}}
\newcommand{\eeq}{\end{equation}}

\def\bk{{\mbox{\boldmath$k$}}}
\def\bq{{\mbox{\boldmath$Q$}}}

\def\bp{{\mbox{\boldmath$p$}}}

\newcommand{\CM}{{\cal M}}

\newcommand{\CO}{{\cal O}}

\begin{document}

\title{On isovector meson exchange currents in the Bethe-Salpeter
  approach}
\author{S.G. Bondarenko, V.V. Burov}
\address{Bogoliubov Laboratory of Theoretical Physics, JINR Dubna,
141980 Russia}
\author{M. Beyer}
\address{Department of Physics, Rostock
University, 18051 Rostock, Germany}
\author{S.M. Dorkin}
\address{Far Eastern State University Vladivostok, 690000 Russia}

\maketitle

\begin{abstract}
 We investigate the nonrelativistic reduction of the Bethe-Salpeter
  amplitude for the deuteron electrodisintegration near threshold
  energies. To this end, two assumptions have been used in the
  calculations: 1) the static approximation and 2) the one iteration
  approximation. Within these assumptions it is possible to recover the
  nonrelativistic result including a systematic extension to
  relativistic corrections. We find that the so-called pair current
  term can be constructed from the $P$-wave contribution of the
  deuteron Bethe-Salpeter amplitude. The form factor that enters into
  the calculation of the pair current is constrained by the
  manifestly gauge independent matrix elements.
\end{abstract}


\section{Introduction}

In recent years, the discussion of relativistic issues in the reactions
involving the deuteron has become more and more important. Some well
known examples for a clear experimental evidence have been found in
the electromagnetic disintegration~\cite{cam82,sch92}. This results
are supported by a recent analysis of polarization observables of the
deuteron in a relativistic framework~\cite{lykasov,kaptari}. Also, the
persistent difference between theory and experiment in polarization
observables of proton deuteron scattering have been argued to be of
relativistic origin~\cite{baldin}. It is expected that the
interpretation of new experimental results, e.g., from TJNAF, also calls
for a relativistic treatment.

Meanwhile the number of theoretical investigations devoted to this
question and to tackle the relativistic aspects of few body systems is
increasing. One may basically recognize two lines: on one hand the
majority of approaches uses a nonrelativistic scheme of calculation
and take into account some leading order corrections, such as
nonnucleonic degrees of freedom like mesonic exchange currents,
$\Delta\Delta$-configurations, $NN^{\star}$-states and more. For a
review see, e.g.,
Refs.~\cite{lykasov,kaptari,burov,shebeko,friar80,adam95} which also
contain an large number of references.  Note, however, that the
problem of consistency arises as has been recently pointed out again,
e.g., by Ref.~\cite{ritz96} in the context of electrodisintegration.
The second line recognizes the importance of
covariance~\cite{kei91,dev93,tjon95}. 

We follow the Bethe-Salpeter approach which is covariant by
construction and properly defined~\cite{lurie,IS,nakanishi}.  Based on
the field theory of particles this approach is also consistent and
automatically takes into account all relativistic effects in the two
nucleon system.  Because of the covariance of the Bethe-Salpeter
equation the amplitudes (or vertex functions) may be written in any
reference system, and one may chose a convenient one.  However,
practical applications are hampered by the difficulty of finding
solutions of the Bethe-Salpeter equation, even in ladder
approximation.  Nevertheless, the activity on solving the
Bethe-Salpeter equation has increased in recent
years~\cite{umn94,nie96}, and by now parameterizations of the
Bethe-Salpeter amplitudes in Euclidean space are
available~\cite{umn96}. Also, useful results have been achieved by a
reduction of the Bethe-Salpeter equation to a three dimensional one.

Here, we examine the nonrelativistic reduction of the Bethe-Salpeter
approach to compare with the results achieved by nonrelativistic
calculations. Since the problem has not been solved so far for the
general case, we restrict ourself to a special process that is the
disintegration of the deuteron.To
this end we consider the threshold region only that exhibits already
the main features to be discussed. Also, this region is dominated by a
single transition amplitude to the $^1S_0$ final state (see, e.g.,
Ref.~\cite{are94}) thus leading to some technical but not substantial
simplifications. The extension to other partial amplitudes is
straightforward but tedious, however beyond the scope of our present
approach.

Indeed, the deuteron disintegration close to  threshold energies
has been a key reaction to establish subnucleonic degrees of freedom
in  nuclei~\cite{edexp} and it attracts continuous attention from
theoretical and experimental sides. It has been an excellent paradigm
to examine nonnucleonic degrees of freedom and relativistic effects.
It is well known that the nonrelativistic impulse
approximation fails to describe the double differential cross section
at momentum transfer squared $-q^2 > 9$ fm$^{-2}$. The experimental
data do not indicate the deep minimum present in the
calculations~\cite{edtheor}.  To fill this minimum mesonic exchange
currents have been introduced.  The contributions of $\pi$-,
$\rho$-currents, and $\Delta \Delta$- configurations allow one to get a
satisfactory agreement with the data ~\cite{mathiot}.  The question of a
consistent inclusion of all relativistic corrections (at least for the
pion sector) has recently been addressed by Ref.~\cite{ritz96} and
supports the above statement.  Nevertheless, some conceptional
problems of the theory still remain open questions. Among them is the
problem of gauge invariance and the question of the nucleon form
factor to be chosen in the exchange currents
contributions~\cite{f1ge,arenh,sch91}.

By now some relativistic methods identify corrections like meson
exchange currents in the relativistic  impulse
approximation. It has been shown that in the framework of the
light-cone approach the so-called extra components $f_5$ of the
deuteron wave function, and $g_2$ of the $^1S_0$-wave function
introduced in Ref.~\cite{karmanov} give an expression in the
nonrelativistic limit, that analytically equals to the contribution of
the pair current~\cite{karmanov}. In this context the one iteration
assumption that will be explained below for solving the dynamical
equation is substantial. Another important result in this context is
the calculation of the static electromagnetic properties within the
Bethe-Salpeter approach. It follows that the contribution of the
$P$-states to the magnetic moment of the deuteron is numerically close
to the contribution of mesonic currents and agrees in sign~\cite{kkb}.
It has already been noticed earlier in the context of covariant
reductions of the Bethe-Salpeter equation to three dimensional ones
that negative energy components in the wave function are responsible
for pair current type contributions, see, e.g.,
Ref.~\cite{fewbody1997}. Here we show that the suggested reduction
procedure leads to the same {\em analytical} structure as the
nonrelativistic pair current correction. 

The paper is organized as follows. Section~\ref{sec:nonrel} contains
the general formulas for the deuteron disintegration amplitude in
different representations. The static approximation is also discussed.
The one iteration approximation is formulated in
Section~\ref{sec:oneit} where we give reasons for its validity in
simple models. The derivation of the main correction is done in the
Section~\ref{sec:emc}.  We present and discuss our results in
Section~\ref{sec:results}, summarize and conclude in
Section~\ref{sec:conclusion}.

\section{Nonrelativistic reduction of the electromagnetic current
  matrix element}
\label{sec:nonrel}

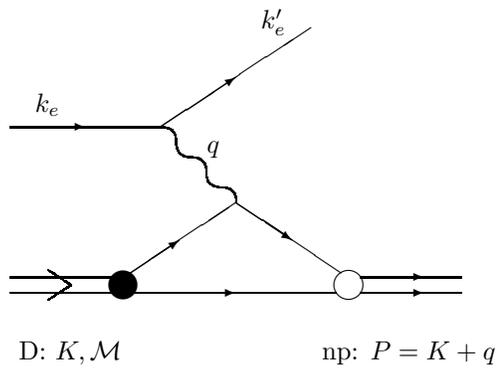
\begin{figure}
\unitlength=1mm
\special{em:linewidth 0.4pt}
\linethickness{0.4pt}
\begin{picture}(50.00,70.00)

\put(20.00,60.00){\line(1,0){20.00}}
\put(40.00,60.00){\line(3,2){20.00}}
\put(20.00,60.00){\vector(1,0){10}}
\put(40.00,60.00){\vector(3,2){10}}
\put(35.00,40.00){\line(3,2){15.00}}
\put(35.00,40.00){\vector(3,2){7.5}}
\put(50.00,50.00){\line(3,-2){14.00}}
\put(50.00,50.00){\vector(3,-2){7.5}}
\put(35.00,38.00){\line(1,0){28.3}}
\put(35.00,38.00){\vector(1,0){15.00}}
\put(20.00,40.00){\line(1,0){15.00}}
\put(20.00,38.00){\line(1,0){15.00}}
\bezier{40}(25.00,41.00)(28.00,39.00)(28.00,39.00)
\bezier{40}(25.00,37.00)(28.00,39.00)(28.00,39.00)
\put(35.00,39.00){\circle*{4.00}}
\put(65.00,39.00){\circle{4.00}}
\put(66.6,40.00){\vector(1,0){8.4}}
\put(66.6,38.00){\vector(1,0){8.4}}
\put(75.00,40.00){\line(1,0){5.00}}
\put(75.00,38.00){\line(1,0){5.00}}
\bezier{40}(40.00,60.00)(42.00,60.00)(42.00,58.00)
\bezier{40}(42.00,58.00)(42.00,56.00)(44.00,56.00)
\bezier{40}(44.00,56.00)(46.00,56.00)(46.00,54.00)
\bezier{40}(46.00,54.00)(46.00,52.00)(48.00,52.00)
\bezier{40}(48.00,52.00)(50.00,52.00)(50.00,50.00)
\put(28.00,30.00){\makebox(0,0)[cc]{D: $K,{\cal M}$}}
\put(73.00,30.00){\makebox(0,0)[cc]{np: $P=K+q$}}
\put(47.00,57.00){\makebox(0,0)[cc]{$q$}}
\put(25.00,63.00){\makebox(0,0)[cc]{$k_e$}}
\put(55.00,74.00){\makebox(0,0)[cc]{$k_e^{\prime}$}}
\end{picture}
 \caption{\label{fig:diagram}
Impulse approximation of the deuteron disintegration process. 
   }
\end{figure}

Because of Lorentz covariance, transformation properties under parity
and time reversal as well as current conservation, the general structure of
the transition matrix element for the $1^+\rightarrow 0^+$ transition
current near threshold can be written as~\cite{bbbd:elec}
\begin{eqnarray}
\langle np (^1S_0) | j_{\mu} | D {\cal M} \rangle =
i\epsilon_{\mu\alpha\beta\gamma}\,\xi^{\alpha}_{\cal M}\,
q^{\beta}\,K^{\gamma}\,V(s,q^2),
\label{formc}
\end{eqnarray}
where $\xi_{\cal M}$ is the deuteron polarization four vector with projection
$\CM$, $K$ is the deuteron
four momentum, $q=P-K$ is the four momentum transfer, $s=P^2=(K+q)^2 \approx
4m^2$ is the squared total momentum of the $np$-pair, see
Fig.~\ref{fig:diagram}. The scalar
function $V(s,q^2)$ may be splitted into isovector Dirac ($F^V_1$) and Pauli
($F^V_2$) contributions
\begin{eqnarray}
V(s,q^2) = V_1(s,q^2) F^{(V)}_1(q^2) + V_2(s,q^2) F^{(V)}_2(q^2).
\label{eqn:V12}
\end{eqnarray}
The final state of the $np$-pair is supposed to be
in a $^1S_0$-state. This is supported by nonrelativistic
calculations~\cite{mathiot}. Eq.~(\ref{formc}) is valid in
the any reference system and the functions $V_{1,2}$ depend on the
scalars $s$ and $q^2$ only. Current conservation is fulfilled because
the  tensor $\epsilon_{\mu\alpha\beta\gamma}$ that
appears in Eq.~(\ref{formc}) is antisymmetric.

To calculate the functions $V_{1,2}$ from the underlying dynamics in
the Bethe-Salpeter approach, two representations of the Bethe-Salpeter
amplitude may be used in concrete calculations: 1) the covariant
representation and 2) the partial-wave representation.  Using the {\em
  covariant representation} for the initial and final states allows us
to directly read off the functions $V_{1,2}(s,q^2)$ from the proper
current matrix elements.  This has been done explicitly in
Ref.~\cite{bbbd:elec}.  However, to study the relation to
the nonrelativistic expressions it is more suitable to give the
formulas for $V_{1,2}(s,q^2)$  in terms of a {\em partial-wave
representation} of the vertex functions.  Hence, the expressions for
$V_{1,2}$ can be written as
\begin{equation}
V_{1,2}(s,q^2)=
i \int d^4k
\sum_{\alpha,\beta}\ G_{\alpha}(p_0,|\bp|)\
{\cal O}_{\alpha\beta}\ g_{\beta}(k_0,|\bk|),
\label{a2}
\end{equation}
where the partial vertex functions $G_{1-4}(p_0,|\bp|)$ represent the
states $^1S_0^{++}$, $^1S_0^{--}$, $^3P_0^{e}$, $^3P_0^{o}$ of the
$np$-pair, and $g_{1-8}(k_0,|\bk|)$ denote the states $^3S_1^{++}$,
$^3S_1^{--}$, $^3D_1^{++}$, $^3D_1^{--}$, $^3P_1^{e}$, $^3P_1^{o}$,
$^1P_1^{e}$, $^1P_1^{o}$ of the deuteron. We use spectroscopic
notation and ($\pm$, $e/o$) as $\rho$-spin quantum
numbers~\cite{kubis}.  In general, the lengthy functions ${\cal
  O}_{\alpha\beta}$ depend on Lorentz scalars and the explicit
expressions are omitted here. We will specify them below after having
introduced appropriate approximations.  The expressions for $p_0$ and
$|\bp|$ are given in a formally covariant way, viz.
\begin{equation}
p_0 = \frac{(K+q,k+q/2)}{\sqrt{s}},\
|\bp|=\left(\frac{(K+q,k+q/2)^2}{s}-(k+q/2)^2\right)^{1/2},
\label{a3}
\end{equation}
where $(\,,)$ denotes the Lorentz scalar product.

We note that through the use of Bethe-Salpeter vertex functions the
denominators of ${\cal O}_{\cal \alpha\beta}$ appearing in
Eq.~(\ref{a2}) contain products of $(\pm M/2 \pm k_0 \pm E_k)$ and
$(\pm \sqrt{s}/2 \pm p_0 \pm E_p)$, where $E_p=\sqrt{\bp^2+m^2}$, etc.
that stem from the nucleon propagators (see Fig.~\ref{fig:diagram}).
We evaluate the integrals in the laboratory system (deuteron at rest).
At threshold, because of the small deuteron binding energy, it is
possible to utilize the static approximation~\cite{ZT} that preserves
the analytical structure of Eq.~(\ref{a2}) through the following
equations
\begin{equation}
p_0 = k_0, \quad |\bp| = | \bk + \frac{\bq}{2} | =
\left(\bk^2 + \frac{\bq^2}{4} + |\bk| |\bq |{x}\right)^{1/2}.
\label{eqn:static}
\end{equation}
Here and in the following we use ($\hat{\bf a}={\bf a}/|{\bf a}|$) 
\begin{eqnarray}
Q&=&|\bq|=((M^2+s-q^2)^2-4M^2s)^{1/2 }/(2M),\ 
{x}=\hat\bk\cdot\hat\bq, \nonumber\\
q&=&(\omega,\bq),\ \omega=(s-M^2-q^2)/(2M).
\label{qaoth}
\end{eqnarray}

We illustrate the approximation by looking closer to the expressions
involving the $(++)$ states.  The  full
denominator of the integrand in
Eq.~(\ref{a2}) leads to a complicated pole structure and reads explicitly
\begin{eqnarray}
\left(\frac{s}{4} -(k_0+\frac {\omega}{2})^2+\bp^2-\sqrt{s}
\left(\frac {Q^2}{s}(k_0+\frac{\omega}{2})^2+E_p^2\right)^{1/2}
-i\epsilon\right)
 \left(\frac{M}{2}+k_0-E_k+i\epsilon\right).
\label{eq8}
\end{eqnarray}
Using Eq.~(\ref{eqn:static}), then leads
to a simple pole expression for the integrand involving the denominator
\begin{eqnarray}
\left(\frac {\sqrt{s}}{2}+k_0-E_p+i\epsilon\right) 
\left(\frac {\sqrt{s}}{2}-k_0-E_p+i\epsilon\right)
\left(\frac {M}{2}+k_0-E_k+i\epsilon\right).
\end{eqnarray}
Thus, static approximation means in particular that $\omega=0$ (no
retardation) and the Lorentz boost transformation of the $np$-pair
vertex functions is neglected.  Going beyond the static approximation
can be achieved by expanding the full expression in terms of $\omega/M $
and $Q^2/s$  that lead to additive corrections.  

The integration on $k_0$ can now be performed by choosing a proper
integration contour and specifying the corresponding poles, e.g.,
closing the upper half plane leads to poles for $k_0$ at $\bar
p_0=\sqrt{s}/2-E_p$. The vertex functions are then evaluated at $(\bar
p_0, k)$.  Since in the reaction under consideration $s\approx 4m^2
\approx M^2$, we expand the vertex functions near $\bar k_0=M/2-E_k$
for $g_1$ and $G_1$ respectively that allows us to derive analytical
expressions in the one iteration approximation as will be shown in the
next section. The analogous procedure holds for the other partial
vertex functions. With this choice one the nucleons in the deuteron is
taken on-shell.

Alternatively, it would have been possible to expand the functions,
e.g., near $k_0=0$. However, the above advocated approach is more
suitable, since it leads to an equation that allows for an analytical
of $k_0$.

In Eq.~(\ref{a2}) we now perform the $k_0$ integration as explained
above. The angular integration is simplified by taking $\bq$ along the
$z$ axis and replace $x$ given in Eq.~(\ref{qaoth}) via
\begin{equation}
x=\frac{{k^{\prime}}^2-k^2-Q^2/4}{kQ},\quad
dx = \frac{2k^{\prime}}{kQ}\,dk^{\prime},
\label{ttt}
\end{equation}
where $k=|\bk|$.  Finally, to compare Eq.~(\ref{a2}) with the
nonrelativistic limit we introduce a $k/m$-expansion excluding order
$\CO (k^2/m^2)$ (i.e. $E_k=m+\CO (k^2/m^2)=E_p$). The resulting
structure functions $V_{1,2}(s,q^2)$ are then given by
\begin{eqnarray}
V_1(s,q^2)
&=&
\frac{\pi}{mM}\frac{1}{Q}
\int\limits_{0}^{+\infty}\,k\,dk\,
\int\limits_{|k-Q/2|}^{k+Q/2}\,k^{\prime}\,dk^{\prime}\,
\Biggl\{
\frac{G_1({\bar k_0^{\prime}},k^{\prime})}
{\sqrt{s}-2E_{k^{\prime}}+i\epsilon}\,
\biggl(\frac{g_1({\bar k_0},k)}{M-2E_k}-
\frac{1}{\sqrt{2}}\,\frac{g_3({\bar k_0},k)}{M-2E_k}
\,P_2(x)\biggr)
\nonumber\\
&&-\frac{\sqrt{3}}{2}\,\frac{G_1({\bar k_0^{\prime}},k^{\prime})}
{\sqrt{s}-2E_{k^{\prime}}+i\epsilon}\,
\biggl( g_5({\bar k_0},k)-g_6({\bar k_0},k) \biggr)\,\frac{x}{Q}
\nonumber\\
&&-\frac{\sqrt{2}}{4}
\biggl( -G_3({\bar k_0^{\prime}},k^{\prime})+G_4({\bar
  k_0^{\prime}},k^{\prime})
\biggr)
\nonumber\\
&&
\times\biggl( \frac{g_1({\bar k_0},k)}{M-2E_k}\,
\frac{Q+2kx}{Qk^{\prime}}
-\frac{1}{\sqrt{2}}\,
\frac{g_3({\bar k_0},k)}{M-2E_k}\,
\frac{QP_2(x)+2kx}{Qk^{\prime}}
\biggr)
\Biggr\},
\label{v1nred}\\
V_2(s,q^2)
&=&
\frac{\pi}{mM}\frac{1}{Q} \int\limits_{0}^{+\infty}\,k\,dk\,
\int\limits_{|k-Q/2|}^{k+Q/2}\,k^{\prime}\,dk^{\prime}\,
\Biggl\{
\frac{G_1({\bar k_0^{\prime}},k^{\prime})}
{\sqrt{s}-2E_{k^{\prime}}+i\epsilon}\,
\biggl(\frac{g_1({\bar k_0},k)}{M-2E_k}
-\frac{1}{\sqrt{2}}\,\frac{g_3({\bar k_0},k)}{M-2E_k}
P_2(x)\biggr)
\nonumber\\
&&+\frac{\sqrt{3}}{4}\,
\frac{Q^2}{4m^2}\,
\frac{G_1({\bar k_0^{\prime}},k^{\prime})}
{\sqrt{s}-2E_{k^{\prime}}+i\epsilon}\,
\biggl( g_5({\bar k_0},k)-g_6({\bar k_0},k) \biggr)\,\frac{x}{Q}
\nonumber\\
&&+ \frac{3\sqrt{2}}{16}\;
\frac{Q^2}{4m^2}\,
\biggl( -G_3({\bar k_0^{\prime}},k^{\prime})+
G_4({\bar k_0^{\prime}},k^{\prime}) \biggr)
\nonumber\\
&&\times\biggl( \frac{g_1({\bar k_0},k)}{M-2E_k}\,
\frac{Q+2kx}{Qk^{\prime}}
-\frac{1}{\sqrt{2}}\,\frac{g_3({\bar k_0},k)}{M-2E_k}\,
\frac{Q\,P_2(x)+2kx}{Qk^{\prime}}
\biggr)
\Biggr\}.
\label{v2nred}
\end{eqnarray}
where ${\bar k_0^{\prime}}\equiv {\bar
  p}_0=\sqrt{s}/2-E_{k^{\prime}}$, and ${\bar k_0}=M/2-E_k$, and
$P_2(x)=(3x^2-1)/2$ is the Legendre polynomial.  The functions $g_2$,
$g_4$, $g_7$, $g_8$, $G_2$ disappear in the above expressions after
$k_0$ integration and because of the $k/m$ expansion.  Note, that
within this approximation scheme we are left with the $(++)$ to $(++)$
and $(++)$ to $(e,o)$ transitions only. All other matrix elements,
such as $(--)$ to all and $(e,o)$ to $(e,o)$ cancel, respectively.

We now examine the expressions for $V_{1,2}(s,q^2)$ given in
Eqs.~(\ref{v1nred},\ref{v2nred}) more closely.  To recover the
nonrelativistic result, we neglect the vertex functions $g_5$,
$g_6$, $G_3$, $G_4$ that correspond to the negative $\rho$ spin components (i.e.
do not exist in the nonrelativistic scheme).  If we replace the
functions $g_1$ and $g_3$ by the nonrelativistic $S$ and $D$ wave
functions and $G_1$ by the $^1S_0$ continuum wave function in the
following way
\begin{eqnarray}
\frac{g_1({\bar k_0},k)}
{M-2E_k} &\rightarrow& -\alpha_1\, u(k),
\quad\quad\alpha_1=4\pi\sqrt{2M}
\nonumber\\
\frac{g_3({\bar k_0},k)}{M-2E_k} &\rightarrow& -\alpha_1\,  w(k),
\nonumber\\
\frac{G_1({\bar k_0^{\prime}},k^{\prime})}
{\sqrt{s}-2E_{k^{\prime}}+i\varepsilon}
&\rightarrow & -\alpha_2\,
u_0(k^{\prime}),\quad\quad \alpha_2=\frac{1}{\sqrt{4\pi}}\frac{1}{2\pi},
\label{r2n}
\end{eqnarray}
and insert these for the respective vertex functions into
Eqs.~(\ref{v1nred},\ref{v2nred}) we obtain
\begin{equation}
V^{(0)}(s,q^2) = \frac{\alpha_1\alpha_2\pi}{mM}\frac{1}{Q}\,
G_M^{(V)}(q^2)\,
\int\limits_{0}^{+\infty}\,k\,dk\,
\int\limits_{|k-Q/2|}^{k+Q/2}\,k^{\prime}\,dk^{\prime}\,
u_0(k^{\prime})
\biggl( u(k)-\frac{1}{\sqrt{2}}\,w(k)\,P_2(x) \biggr).
\label{vnrnP}
\end{equation}
where we have introduced the magnetic isovector form factor
$G_M^{(V)}=F_1^{(V)}+F_2^{(V)}$. In co-ordinate space the respective
integral is achieved using the following transformations,
for the deuteron states ($w_0\equiv
u$, $w_2\equiv w$, $\ell = 0,2$)
\begin{equation}
w_\ell(k)=\int\limits_{0}^{+\infty}\,r\,dr\,
w_\ell(r)\,j_\ell(kr),\quad\quad
\frac{w_\ell(r)}{r} = \frac{2}{\pi}
\int\limits_{0}^{+\infty}\,k^2\,dk\,w_\ell(k)\,j_\ell(kr),
\label{nr2k}
\end{equation}
where $j_\ell(x)$ is spherical Bessel function, and for the scattering
state 
\begin{eqnarray}
u_0(k) =
\frac{2}{\pi}\,\int\limits_{0}^{+\infty}\,r\,dr\,u_0(r)\,j_0(kr),\quad\quad
\frac{u_0(r)}{r} =
\int\limits_{0}^{+\infty}\,k^2\,dk\,u_0(k)\,j_0(kr).
\nonumber\end{eqnarray}
The resulting expression is
\begin{equation}
V^{(0)}(s,q^2)=\frac{\alpha_1\alpha_2\pi}{mM}\,
G_M^{(V)}(q^2)\,
\int\limits_{0}^{+\infty}\,dr\,
u_0(r)
\biggl( u(r)\,j_0(Qr/2)-\frac{1}{\sqrt{2}}\,w(r)\,j_2(Qr/2) \biggr),
\label{vnrn}
\end{equation}
This result reflects the so-called nonrelativistic impulse
approximation and represents the lowest order nonrelativistic
expansion of the transition form factors given in
Eqs.~(\ref{v1nred},\ref{v2nred}).

\section{One iteration approximation}
\label{sec:oneit}

\subsection{Deuteron channel}
The Bethe-Salpeter equation is commonly solved by iterations. After
angular decomposition we are left with an integral equation for the
radial parts of the vertex function $g_{\alpha}(p_0,p)$. They are
connected to the partial amplitudes $\psi_{\beta}(k_0,k)$ which are
used here for simplicity on the right hand side of the following
equation by amputation~\cite{kubis}, e.g., for the $(++)$ components
the relation is given by
\begin{equation}
\psi_\alpha = \frac{g_\alpha}
{(M/2 + k_0 -E_k)(M/2-k_0-E_k)},\qquad \alpha=1,3.
\end{equation}
 The integral equation read
\begin{eqnarray}
g_{\alpha}(p_0,p)=\,\sum_{M}\frac{g_{MNN}^2}{4\pi}\,\frac{-i}{\pi^2}
\int\limits_{-\infty}^{+\infty}\,dk_0\,\int\limits_{0}^{+\infty}\,
\frac{1}{4E_kE_p}\,\frac{k}{p}\,dk\,\sum_{\beta}
V^{(M)}_{\alpha\beta}(p_0,p;k_0,k)
\psi_{\beta}(k_0,k),
\label{bseq1}
\end{eqnarray}
where $M$ denotes the type of exchanged meson, $g_{MNN}$ its coupling
constant and $V^{(M)}_{\alpha\beta}$ is the transition matrix element
between the states $\alpha$ and $\beta$ given below. To get fast
convergence to the solution of this equation one needs a good educated
guess for the initial vertex function $g_{\alpha}(p_0,p)$ respectively
$\psi_{\beta}(k_0,k)$.  This guess may be taken from the solution of
the equivalent nonrelativistic problem. After several iterations one
usually gets the exact solution.

The initial vertex functions may be chosen along the lines
given in Ref.~\cite{karmanov}, generalizing Eq.~(\ref{r2n})
\begin{eqnarray}
\psi_{1}(k_0,k) &= &\frac{-\alpha_1\, (M-2E_k)\,
u(k)}{(M/2+k_0-E_k+i\epsilon)(M/2-k_0-E_k+i\epsilon)}\;,
\label{eqn:psi1}\\
\psi_{3}(k_0,k) &=& \frac{-\alpha_1\, (M-2E_k)\,
w(k)}{(M/2+k_0-E_k+i\epsilon)(M/2-k_0-E_k+i\epsilon)}\;,
\label{eqn:psi3}
\end{eqnarray}
 All other wave functions are assumed to
vanish in this order. Note, that the numerators of
Eqs.(\ref{eqn:psi1},\ref{eqn:psi3}) are the nonrelativistic vertex
functions. 

The ansatz given in Eqs.~(\ref{eqn:psi1}) and (\ref{eqn:psi3}) has
been numerically verified for the deuteron consisting of two spinor
nucleons that interact via the exchange of scalar mesons with mass
$\mu = 0.2\,m$.  This interaction leads to a bound state for both,
nonrelativistic and relativistic cases. In the {\em nonrelativistic}
case and use of the Schr\"odinger equation the kernel is of the form
\begin{equation}
V(\bp-\bk)=\frac{g_{NR}^2}{4\pi}\frac{1}{(\bp-\bk)^2+\mu^2},
\end{equation}
where the coupling constant $g_{NR}=8.37$ leads to a binding energy of
$\epsilon= 2.23$ MeV. The resulting wave function of the deuteron may
be parameterized in the following form,
\begin{equation}
u(p)=N\left(\frac{1}{p^2+\lambda^2}
-\frac{0.40}{p^2+(\lambda+\mu)^2}-\frac{0.60}{p^2+(\lambda+2\mu)^2}\right),
\label{wfnr}
\end{equation}
with $\lambda=\sqrt{m \epsilon}$ and a normalization constant $N$.
Note, that for the scalar exchange model the $D$-wave function is
zero, $w(p)=0$.  These functions are then substituted as an educated
guess into the system of equation (\ref{bseq1}) using the assumption
of Eqs.~(\ref{eqn:psi1},\ref{eqn:psi3}).  Eq.~(\ref{bseq1}) is then
solved in Euclidean space ($p_0\rightarrow ip_4$) using the ansatz
amplitude Eq.~(\ref{wfnr}) and the full kernel $V(p_0,p;k_0,k)$. 
\begin{equation}
V(p_0,p;k_0,k) = \frac{g^2}{4\pi} \frac{1}{(k-p)^2-\mu^2}
{\cal I}\otimes{\cal I}
\end{equation}
To achieve the same deuteron binding energy as before, now using the
Bethe-Salpeter equation (\ref{bseq1}) the coupling constant $g$ of the
scalar interaction needs to be readjusted to get the same deuteron
binding energy, viz.  $g=17.00$.

In this framework we compare the full vertex function of the exact
solution of Eq.~(\ref{bseq1}) to the result achieved by one iteration
only but using the nonrelativistic solution as an educated ansatz. This
will be called {\em one iteration approximation} in the
following.  As a demonstration for the quality of the approximation
we show two vertex functions, for the $^3S_1^{++}$- and
$^3P_1^o$-channels in Fig.~\ref{fig:compare}. The upper two surfaces
displayed correspond to the exact result and the one iteration
approximation of the $^3S_1^{++}$-component. The difference between
the two is rather small. In fact, in the relevant region $p=0\dots0.5$
GeV and $p_4=-0.1\dots0.1$ GeV, where the amplitude has its major
contribution to the normalization and to physical processes, the
difference does not exceed 5\%. The exact and the one iteration
approximation for the $^3P_1^o$-channel that is relevant for the
subsequent discussion of the electromagnetic properties are displayed
as the lower two surfaces.  We conclude that for the scalar model the
shape of the exact solutions are already reproduced by one
iteration.  From that we argue that the one iteration approximation
is a reasonable first step that may be utilized to solve the more
complicated problem with realistic potentials.

We now turn back to the discussion of the realistic deuteron
implementing the full kernel with all spin dependences. By using
Eqs.~(\ref{bseq1}) along with Eqs.~(\ref{eqn:psi1},\ref{eqn:psi3}) the
expressions for the $^3P_1^e(^3P_1^o)$-channel, i.e. for the functions
$g_5(g_6)$ are now deduced analytically.  If we consider the
$\pi$-exchange kernel only and use the above given notation for the
states ($1\dots 8$) then
\begin{eqnarray}
V_{51}(p_0,p;k_0,k) &= &V_{53}(p_0,p;k_0,k) = 0,
\\
V_{61}(p_0,p;k_0,k) &=& \frac{2}{\sqrt3}
\left(\frac{1}{3}\,p\,(E_k-m)\,Q_2(z)
+m\,k\,Q_1(z)-\frac{1}{3}\,p\,(E_k+2m)Q_0(z)\right),
\\
V_{63}(p_0,p;k_0,k) &= &\frac{\sqrt{2}}{\sqrt{3}}
\left(\frac{1}{3}\,p\,
(2E_k+m)\,Q_2(z)-m\,k\,Q_1(z)-\frac{2}{3}\,p\,(E_k-m)Q_0(z)\right),
\end{eqnarray}
with Legendre functions of the second kind $Q_\ell(z)$, and
$z={(p^2+k^2+\mu_{\pi}^2-(k_0-p_0)^2)/2pk}$.
The $k_0$ integration can be performed yielding (e.g. for the
pionic exchange kernel) 
\begin{eqnarray}
g_{\alpha}({\bar p_0},p) = \frac{\alpha_1}{2\pi m^2}\,
\frac{g_{\pi NN}^2}{4\pi}\,
\int\limits_{0}^{+\infty}\,
\frac{k}{p}\,dk\,\left(
\bar V^{(\pi)}_{\alpha\, 1}(p,k) u(k) +
\bar V^{(\pi)}_{\alpha\, 3}(p,k) w(k) \right),\quad
\alpha = 5,6.
\label{bseq2}
\end{eqnarray}
The functions $\bar V^{(\pi)}_{\alpha\, (1,3)}(p,k)$ are obtained from
$V^{(\pi)}_{\alpha\, (1,3)}(p_0,p;k_0,k)$ by treating $p_0$ and $k_0$
as explained in Sect.~\ref{sec:nonrel}.  The explicit expressions for
$\bar V$ ($5,6\leftarrow 1,3$) are given by:
\begin{eqnarray}
\bar V_{51}(p,k) &=&\bar  V_{53}(p,k) = 0,
\label{pot1}\\
\bar V_{61}(p,k)& =& \frac{2m}{\sqrt{3}}(k\,Q_1(\bar z)-p\,Q_0(\bar z)),
\label{pot2}\\
\bar V_{63}(p,k)& =& \frac{\sqrt{2}m}{\sqrt{3}}(p\,Q_2(\bar z)-k\,Q_1(\bar z)),
\label{pot3}
\end{eqnarray}
with $\bar z={(p^2+k^2+\mu_{\pi}^2)/2pk}$.
To simplify the notation the isospin factors  have been
omitted in $\bar V$ and Eq.~(\ref{bseq2}). The extension to other
mesons of a one boson exchange kernel is straight forward.

Now using Eqs.~(\ref{bseq2},\ref{pot1}), the transformation
Eq.~(\ref{nr2k}) and the integral representations of the Legendre
functions $Q_\ell(z)=\,2pk\,\int\limits_{0}^{+\infty}\,
e^{-\mu_{\pi}r}\,r\,dr\,j_\ell(kr)\, j_l(pr)$ we obtain following
results for the vertex functions:
\begin{eqnarray}
g_5({\bar k_0},k)&=&0,
\label{find}\\
g_6({\bar k_0},k)&=&-(-3)\,\frac{\alpha_1}{\sqrt{3}m}\,
\frac{g_{\pi NN}^2}{4\pi}\,
\int\limits_{0}^{+\infty}\,dr\,\frac{e^{-\mu_{\pi}r}}{r}\,(1+\mu_{\pi}r)
\biggl(u(r)+\frac{1}{\sqrt{2}}w(r)\biggr)\,j_1(kr),
\label{find1}
\end{eqnarray}
where the isospin factor $(-3)$ has now been shown explicitly. This is
the analytical structure given after the first iteration.

\subsection{$^1S_0$-channel}

The inhomogeneous Bethe-Salpeter equation  for the amplitudes
in the $^1S_0$-channel reads
\begin{eqnarray}
\phi_{\alpha}(p_0,p)&=&\phi_{\alpha}^{(0)}(p_0,p)
 +\sum_{M}\frac{g_{MNN}^2}{4\pi}\,\frac{-i}{\pi^2}
\int\limits_{-\infty}^{+\infty}\,dk_0\,\int\limits_{0}^{+\infty}\,
\frac{1}{4E_kE_p}\,\frac{k}{p}\,dk\,\nonumber\\
&&\times\sum_{\beta\gamma}
S_{\alpha\beta}(p_0,p;s)
V^{(M)}_{\beta\gamma}(p_0,p;k_0,k)\phi_{\gamma}(k_0,k).
\end{eqnarray}
Here $\phi_{\alpha}^{(0)}$ denotes the plane-wave function,
\begin{eqnarray}
\phi_{\alpha}^{(0)}(p_0,p)=\frac{1}{\sqrt{4\pi}}\;\delta_{\alpha 1}\;
\delta(p_0)\;\frac{1}{p^2}\;\delta(p-p^*),
\end{eqnarray}
where $p^*$ is the on-energy-shell momentum given by
$p^*=|{\bp}^*|=\sqrt{s/4-m^2}$.

For the subsequent discussion it is more convenient to split the
system of equations into the following form:
\begin{eqnarray}
\phi_{1}(p_0,p)&=&\phi_{1}^{(0)}(p_0,p)
+\sum_{M}\frac{g_{MNN}^2}{4\pi}\,\frac{-i}{\pi^2}
\int\limits_{-\infty}^{+\infty}\,dk_0\,\int\limits_{0}^{+\infty}\,
\frac{1}{4E_kE_p}\,\frac{k}{p}\,dk\,
\nonumber\\
&& \times \sum_{\beta} V^{(M)}_{1\beta}(p_0,p;k_0,k)
\frac{\phi_{\beta}(k_0,k)}{(\sqrt{s}/2+p_0-E_p+i\epsilon)
(\sqrt{s}/2-p_0-E_p+i\epsilon)},\\
G_{\alpha^{\prime}}(p_0,p) &=& \sum_{M}
\frac{g_{MNN}^2}{4\pi}\,\frac{-i}{\pi^2}
\int\limits_{-\infty}^{+\infty}\,dk_0\,\int\limits_{0}^{+\infty}\,
\frac{1}{4E_kE_p}\,\frac{k}{p}\,dk\,
\nonumber\\
&& \times\sum_{\beta}
V^{(M)}_{\alpha^{\prime}\beta}(p_0,p;k_0,k)\phi_{\beta}(k_0,k),
\qquad \qquad \alpha^{\prime}\neq 1=^1\!S_0^{++}.
\end{eqnarray}
Since the full vertex functions are related to
the full amplitudes by amputation, the partial vertex functions
$G_{\alpha}(p_0,p)$ are connected to the partial amplitudes
$\phi_{\alpha}(p_0,p)$ via simple relations~\cite{kubis}.  As in
the case of the deuteron channel we consider the one iteration
approximation. To do so we chose a similar expression for the initial
amplitudes as before to connect to the nonrelativistic solution,
\begin{eqnarray}
\phi_{1}(k_0,k^{\prime}) = \frac{-\alpha_2\, (\sqrt{s}-2E_{k^{\prime}})\,
u_0(k)}{(\sqrt{s}/2+k_0-E_{k^{\prime}}+i\epsilon)
(\sqrt{s}/2-k_0-E_{k^{\prime}}+i\epsilon)}.
\label{eqn:ph1}
\end{eqnarray}
Here $u_0(k)$ is the nonrelativistic continuum wave function in
$^1S_0$ channel given by
\begin{eqnarray}
u_0(k)=\frac{1}{k^2}\;\delta(k-p^*)
+\frac{m\;t(k,p^{*};E_{p^{*}})}{{p^{*}}^{2}-k^2+i\epsilon},
\end{eqnarray}
and $t(k,p^{*};E_{p^{*}})$ is nonrelativistic half-off-shell
$t$-matrix for the $^1S_0$ channel normalized through  the condition
\begin{eqnarray}
t(E_{p^*}) \equiv
t(p^*,p^*;E_{p^*})=-\frac{2}{\pi}\;\frac{1}{m\;p^*}\;
\sin{\delta_0}\;e^{i\delta_0},
\end{eqnarray}
where $\delta_0$ is the phase shift, and $E_{p^*}={p^*}^2/m$.

Analogously to the deuteron case, we finally arrive at the first order
corrections to the amplitudes in the $^1S_0$ channel,
\begin{eqnarray}
G_3({\bar k_0},k^{\prime})&=&
-(+1)\,\frac{\alpha_2\sqrt{2}}{\pi m}\,\frac{g_{\pi NN}^2}{4\pi}\,
\int\limits_{0}^{+\infty}\,dr\,\frac{e^{-\mu_{\pi}r}}{r}\,(1+\mu_{\pi}r)
u_0(r)\,j_1(k^{\prime}r),
\label{finp}\\
G_4({\bar k_0},k^{\prime})&=&0,
\label{finp1}
\end{eqnarray}
where $(+1)$ is the isospin factor.

We have shown in this section that choosing proper zero approximation
wave function (i.e. the nonrelativistic ones) after one interaction
additional partial amplitudes arise through the Bethe-Salpeter
equation. They are connected to the interaction kernel and in
electromagnetic processes give rise to the so called pair current
correction as will be shown in the next section.

\section{The structure of the electromagnetic current matrix element}
\label{sec:emc}

We are now in the position to turn to the first order corrections to
$V^{(0)}(s,q^2)$ given in Eq.~(\ref{vnrn}). To this end we expand the
expressions given in Eqs.~(\ref{eqn:V12}), (\ref{v1nred}) and
(\ref{v2nred}) into a power series of $g_{\pi NN}^2/4\pi$ to extract
the pionic contribution only.  Also we consider the $P$-states
contribution only, i.e. components with one negative $\rho$ spin. The
resulting expression for the transition form factor $V(s,q^2)$ will be
denoted by $V^{(\pi)}(s,q^2)$.  Substituting Eqs.~(\ref{find}) and
(\ref{find1}) as well as Eqs. (\ref{finp}) and (\ref{finp1}) into Eqs.
(\ref{v1nred}) and (\ref{v2nred}) and using the replacements of
Eq.~(\ref{r2n}) then Eq.~(\ref{eqn:V12}) reads
\begin{eqnarray}
V^{(\pi)}(s,q^2)
&=&
\frac{\alpha_1\alpha_2\pi}{2m^2MQ}\;
\frac{g_{\pi NN}^2}{4\pi}
\int\limits_{0}^{+\infty}\,dr \,\frac{e^{-\mu_{\pi}r}}{r}\,(1+\mu_{\pi}r)\,
\int\limits_{0}^{+\infty}\,k\,dk\,
\int\limits_{|k-Q/2|}^{k+Q/2}\,k^{\prime}\,dk^{\prime}
\nonumber\\
&& \times\Biggl\{
(-3)\,\left(F_1^{(V)}(q^2)-\frac{1}{2}\frac{Q^2}{4m^2}\,F_2^{(V)}(q^2)\right)
\,u_0(k^{\prime})
\biggl(u(r)+\frac{1}{\sqrt{2}}w(r)\biggr)\,j_1(kr)\,\frac{x}{Q}
\nonumber\\
&&+
(+1)\,\left(F_1^{(V)}(q^2)-\frac{3}{4}\frac{Q^2}{4m^2}\,F_2^{(V)}(q^2)\right)
\frac{1}{\pi}
u_0(r)\,j_1(k^{\prime}r)\,
\nonumber\\
&&\times\biggl( u(k)\,\frac{Q+2kx}{Qk^{\prime}}
-\frac{1}{\sqrt{2}}\,w(k)\,\frac{QP_2(x)+2kx}{Qk^{\prime}}
\biggr)\Biggr\}.
\end{eqnarray}
The $k^{\prime}$-integration can be solved
analytically.
After doing so, we obtain
\begin{eqnarray}
V^{(\pi)}(s,q^2) &= &\frac{\alpha_1\alpha_2\pi}{2m^2MQ}\;
 \frac{g_{\pi NN}^2}{4\pi}
\int\limits_{0}^{+\infty}\,dr\,\frac{e^{-\mu_{\pi}r}}{r^2}\,(1+\mu_{\pi}r)\,
\nonumber\\
&& \times\Biggl\{
3\left(F_1^{(V)}(q^2)-\frac{1}{2}\frac{Q^2}{4m^2}\,F_2^{(V)}(q^2)\right)
u_0(r)\,\biggl(u(r)+\frac{1}{\sqrt{2}}w(r)\biggr)\,j_1(Qr/2)
\nonumber\\
&&+\left(F_1^{(V)}(q^2)-\frac{3}{4}\frac{Q^2}{4m^2}\,F_2^{(V)}(q^2)\right)
 u_0(r)\,\biggl(u(r)+\frac{1}{\sqrt{2}}w(r)\biggr)\,j_1(Qr/2)
\Biggr\}
\nonumber\\
& =&\frac{2\alpha_1\alpha_2\pi}{m^2MQ}\,
H(q^2)\,\frac{g_{\pi NN}^2}{4\pi}\,
\int\limits_{0}^{+\infty}\,dr\,\frac{e^{-\mu_{\pi}r}}{r^2}\,(1+\mu_{\pi}r)\,
u_0(r)\,\biggl(u(r)+\frac{1}{\sqrt{2}}w(r)\biggr)\,j_1(Qr/2).
\label{ve:final}\end{eqnarray}
Here we have introduced the function
\begin{eqnarray}
H(q^2) = F_1^{(V)}(q^2)-\frac{9}{16}\frac{Q^2}{4m^2}\,
F_2^{(V)}(q^2)
\label{gamma}
\end{eqnarray}
where  $Q$ is given in Eq.~(\ref{qaoth}).

This first order contribution in $g_{\pi NN}^2$ supplements
the lowest order relativistic expansion as given in
Eq.~(\ref{vnrn}) of Sect.~\ref{sec:nonrel}.  We then arrive at the
following expression for the transition form factor:
\begin{eqnarray}
V(s,q^2)
&=&  \frac{\alpha_1\alpha_2\pi}{mM}
\Biggl\{G_M^{(V)}(q^2)\,
\int\limits_{0}^{+\infty}\,dr\,u_0(r)\,\biggl(u(r)\,j_0(Qr/2)-
\frac{1}{\sqrt{2}}w(r)\,j_2(Qr/2)\biggr)
\nonumber\\
&&+H(q^2)\,\frac{2}{mQ}\,\frac{g_{\pi NN}^2}{4\pi}\,
\int\limits_{0}^{+\infty}\,dr\,
\frac{e^{-\mu_{\pi}r}}{r^2}\,(1+\mu_{\pi}r)\,
u_0(r)\,\biggl(u(r)+\frac{1}{\sqrt{2}}w(r)\biggr)\,j_1(Qr/2)\Biggr\}
\label{vnr:final}
\end{eqnarray}
This is our main result. Starting from a fully covariant theory this
result has been achieved using the lowest order relativistic
contribution within the one iteration approximation. The relation
between the nonrelativistic wave functions and the relativistic
amplitudes given in Eqs.~(\ref{eqn:psi1}), (\ref{eqn:psi3}), and
(\ref{eqn:ph1}) allows us to give the final expressions in terms of
nonrelativistic wave functions. The lowest order in the $\pi NN$
coupling constant $g_{\pi NN}^2$ leads to an additional contribution
from iterating the $P$-wave channel once.  Comparing this result to
the one achieved within the nonrelativistic scheme that introduces
mesonic exchange currents, we find find that the first term coincides
analytically with the nonrelativistic impulse approximation
contribution and the second one with the $\pi$-pair current
contribution.

\subsection*{Nonrelativistic impulse approximation and pair current
contribution}

To further illustrate the equivalence we connect our results to the
nonrelativistic formalism describing the break-up process. More
details are given, e.g in Ref.~\cite{NNN6}.  The differential cross
section for the $D \rightarrow {}^{1}S_{0}$ - transition has the form
(nonrelativistic case)
\begin{equation}
\frac{d^{2}\sigma}{d\Omega d\omega} =
\frac{16}{3}\alpha^{2}\frac{k^{\prime 2}_{e}}{Q^2}
\frac{p^* m}{t^{2}}\,\sin^{2}\frac{\theta}{2}\,
((k_{e} + k^\prime_{e})^{2} - 2k_{e}k^\prime_{e}
\cos^{2}\frac{\theta}{2}) \\
|\langle{}^{1}S_{0}\parallel T_{1}^{\rm Mag} \parallel D \rangle| ^{2},
\label{NR}
\end{equation}
where  $q=(\omega,{\bq})$ is the
momentum transfer, $t=-q^2$.  The momentum $p^*$ is related to the
relative energy $E_{p^*}$ of the np system as given before $E_{p^*}
= p^{*\,2}/m$ and the relation between kinematical quantities is
given by $Q = \sqrt{ ((2m + E_{p^*})^{2} - M^{2} + t)^{2}/4M^{2} + t}$
and $k^\prime_{e} = (- \omega + \sqrt{\omega^{2} + t/\sin^{2}\theta/2})/2$,
$\omega = E_{e} - E^\prime_{e}$.

In the general case, the current matrix element is a sum of
nonrelativistic impulse approximation contribution, meson exchange
currents contributions and retardation currents contributions.  Here
we are interested in $\pi$-meson pair current part only and thus we
may write for $\langle{}^{1}S_{0} \parallel T_{1}^{\rm Mag} \parallel
D\rangle$ the following expression:
\begin{eqnarray}
\langle{}^{1}S_{0}  \parallel T_{1}^{\rm Mag} \parallel D\rangle&=&
\langle{}^{1}S_{0}  \parallel T_{1,ia}^{\rm Mag} \parallel D\rangle +
\langle{}^{1}S_{0}  \parallel T_{1,\pi c}^{\rm Mag} \parallel D\rangle
\label{mal}
\end{eqnarray}
where we introduced $T_{1,ia}^{\rm Mag}$, which reflects the impulse
approximation operator and $T_{1,\pi c}^{\rm Mag}$, which is the
$\pi$-meson pair (contact) operator. Expressions for the matrix
elements are given in Ref.~\cite{NNN6}, see Eq.~(11), and may be
compared with Eq.~(\ref{vnr:final}). Differences will be discussed
below.

\section{Results and Discussion}\label{sec:results}

We have shown in the previous section that the nonrelativistic
reduction of the Bethe-Salpeter approach utilizing the one iteration
approximation leads to results that exhibit the same {\em analytical
structure} as the nonrelativistic result plus pair current corrections.
Some details differ as will be discussed below. One may now use the
``exact'' nonrelativistic wave functions to calculate the different
contributions to the cross sections. This has been done here for an
illustration.

The formula for the cross section is given in Eq.~(\ref{NR}).
The dominant M1 transition matrix element (i.e.
$D\rightarrow np(^1S_0)$) of the multipole decomposition given there
directly corresponds to the nonrelativistic reduction of
Eq.~(\ref{formc}) given above. The contribution of the nonrelativistic
impulse approximation given, e.g. in Ref.~\cite{NNN6} coincides with
the formula given in Eq.~(\ref{vnrn}) if $\alpha_1$ and $\alpha_2$
are chosen as in Eq.~(\ref{r2n}).

\begin{figure}[h]
 \hskip 20mm
 \psfig{figure=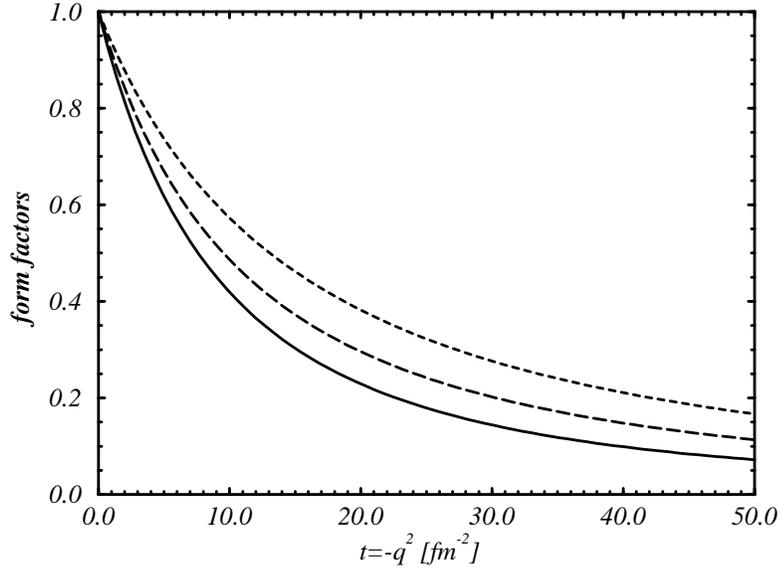,width=0.65\textwidth,angle=270}
 \vskip 5mm
 \caption{
 Nucleon electromagnetic form factors for the nonrelativistic pair
 current contribution as discussed in the text.
 Solid line $F_1^{(V)} (q^2)$, dotted line
 $G_E^{(V)} (q^2)$, and dashed line $H (q^2)$ as given through
 the Bethe-Salpeter approach.}
 \label{fig:ffl}
\end{figure}

The analytical structure of the nonrelativistic pair current
contribution equals to that of the $P$-state contribution derived from
the Bethe-Salpeter approach given in Eq.~(\ref{ve:final}).  We note
that the nonrelativistic pionic pair current contribution
given in Ref.~\cite{NNN6} depends on the nucleon form factor
$F_1^{(V)}(q^2)$. Sometimes also the electric nucleon Sachs form
factor $G^{(V)}_E(q^2)= F^{(V)}_1(q^2) +
\frac{q^2}{4m^2}F^{(V)}_2(q^2)$ has been used (see
Ref.~\cite{suskov}). The Bethe-Salpeter approach yields a different
dependence on the nucleon form factor that is also consistent with
current conservation and given by the function $H(q^2)$ defined in
Eq.~(\ref{gamma}).  As an illustration of the different behavior we
display the form factors $F_1^{(V)}(q^2)$ (solid line),
$G_E^{(V)}(q^2)$ (dotted line), and the function $H(q^2)$ (dashed
line) in Fig.~\ref{fig:ffl}. As a parameterization for the nucleon form
factors we use the one given in Ref.~\cite{hoehler}.  The form factor
$F_1(q^2)$ is larger than the form factor $G_E(q^2)$ and the function
$H(q^2)$ and therefore the respective pionic pair current is expected
to be larger than the one using the other form factors. Since the
function $H(q^2)$ is in the middle of the other ones the respective
contribution of the $P$-states in Bethe-Salpeter approach will be
different from the nonrelativistic calculations normally using
$F_1^{(V)}(q^2)$ or $G_E^{(V)}(q^2)$.

To investigate the influence of the nucleon form factors more closely
we calculated the impulse approximation and pionic pair current
contributions to the differential cross section. The calculation is
performed with the Paris $NN$ potential~\cite{paris-wf} at
$E_{np}=1.5$ MeV and $\theta=155^{\circ}$.

\begin{figure}[h]
 \hskip 20mm
 \psfig{figure=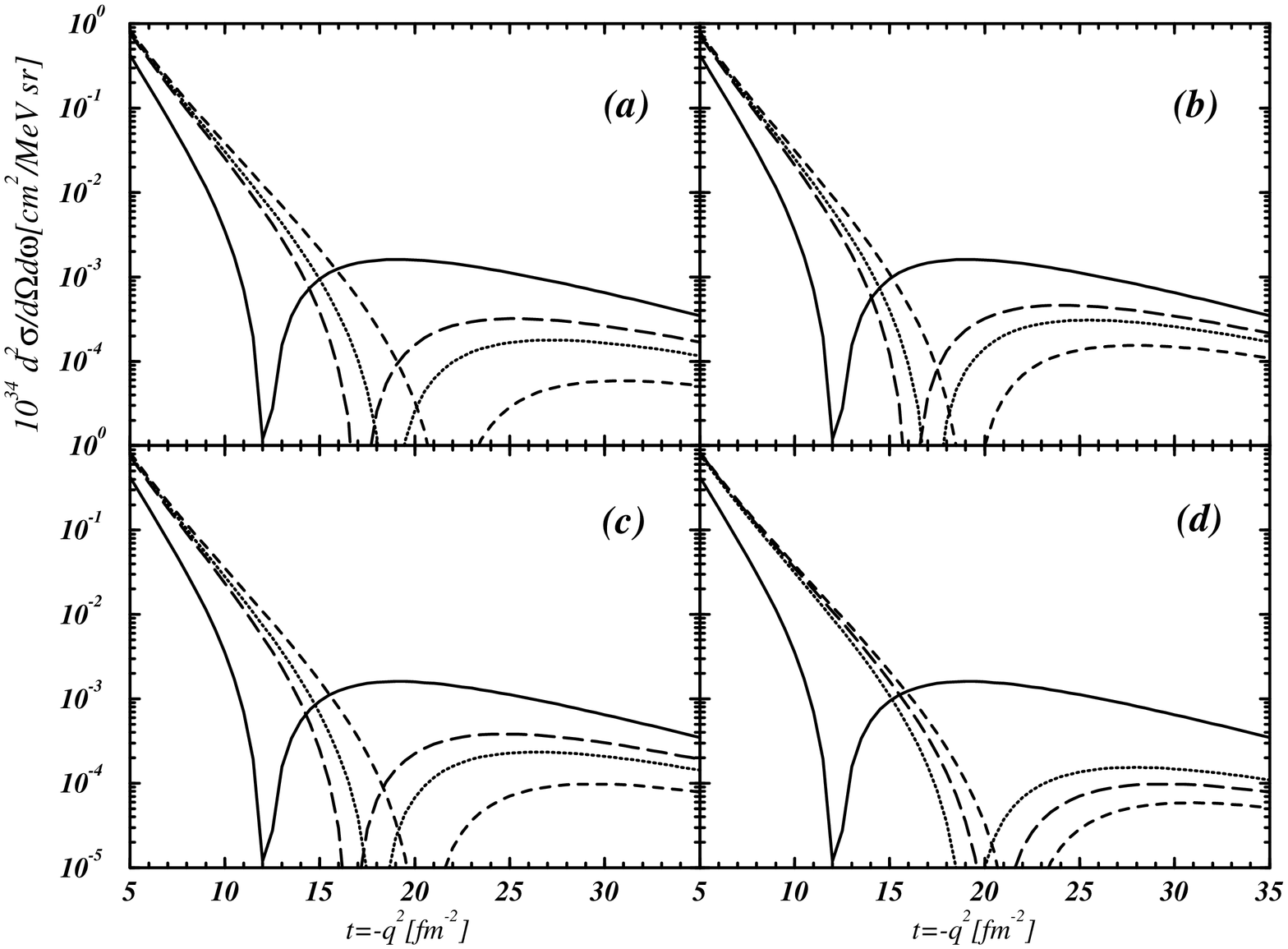,width=\textwidth,angle=270}
 \vspace*{20mm}
 \caption{\label{fig:iapc}
   Impulse approximation (ia, solid lines) and $\pi$-meson
   pair-current contributions (pc) of the differential cross section.
   a-c) different sets of hadronic form factors as explained in the
   text.  Dashed line: ia+pc with $F_1^{(V)} (q^2)$, dotted line:
   ia+pc with $H (q^2)$, long dashed line: ia+pc with $G_E^{(V)}
   (q^2)$.  d) Calculations using $F_1^{(V)} (q^2)$ only but different
   hadronic form factor sets as given in the text. Dashed line:
   ia+pc with set a, dotted line: ia+pc with set b, and long dashed
   line: ia+pc with set c.}
\end{figure}

It is well known that some uncertainty is related to the hadron form
factors. Without dwelling too much on that point we would like to
compare three different form factors to see how this uncertainty
influences the result. Introducing the hadron form factors change
the expressions for the pair current contribution~\cite{pair}.  Three sets
of hadron form factors (for $\pi NN$-vertex) have been utilized and
are shown in Figs.~\ref{fig:iapc} a-c), respectively: a monopole
vertex~\cite{machl87} with a cut-off mass of $\Lambda = 1.25$ GeV (set
a) and $\Lambda = 0.85$ GeV (set b).  In addition a vertex inspired by
a QCD analysis has been used with two parameters given chosen to be
$\Lambda_1 = 0.99$ GeV and $\Lambda_2 = 2.58$ GeV ~\cite{gari83} (set
c).

It is seen from Fig.~\ref{fig:iapc} that there is a strong dependence
of the differential cross section on the nucleon electromagnetic form
factors. It is also seen that the minimum at $t=12$ fm$^{-2}$ in the
impulse approximation contribution is shifted by pionic pair current
to the region $t > 16$ fm$^{-2}$ using $G_E(q^2)$, to $t > 18$
fm$^{-2}$ using the function $H(q^2)$ and to $t > 22$ fm$^{-2}$ using
$F_1^{(V)}(q^2)$ as a form factor. The largest shift in the cross
section is achieved by using $F_1^{(V)}(q^2)$ in the calculations. The
size of the shift and the behavior of the cross section considerably
depends on the set of parameters as well as the type of the $\pi
NN$-vertex used. This is illustrated in Fig.\ref{fig:iapc}d) for
calculations using $F_1^{(V)}(q^2)$ only but for different
parameterizations of the strong vertex.

\section{Summary and Conclusion}
\label{sec:conclusion}

We investigate the electromagnetic current in the framework of the
Bethe-Salpeter approach. The question addressed is the nonrelativistic
reduction of the current transition amplitude. As a specific
application we consider the deuteron disintegration near threshold
energies in some detail. We present a method to reduce the
amplitude to the nonrelativistic one.  We summarize the
basic steps of this method and the main approximations utilized.

To arrive at our result, i.e. to compare the relativistic expressions
to the pair current that appears in the nonrelativistic scheme, two
main assumptions have been introduced: 1) the static approximation and
2) one iteration approximation.  In the static approximation boost
corrections are neglected.  We emphasize that the latter contributions
are not negligible and for the elastic form factors corrections can
achieve 20 \%. A discussion of this contributions in a noncovariant
but consistent relativistic framework may be found in
Ref.~\cite{ritz96}. Although important, a discussion of all these
contributions in not the main issue of our present paper. In fact, we
do not expect any substantial change in our present conclusions, since
all corrections are of additive nature and will be taken into account
as the scheme develops.

The one iteration approximation is to some extent more subtle.
It plays an essential role in connecting the Bethe-Salpeter amplitudes
and the usual nonrelativistic wave functions. This approximation
allows us to express the final result in terms of the usual $S$- and
$D$-wave functions of the deuteron and the $S$-wave function of
$np$-pair.  Although numerically verified for deuteron-like models,
the one iteration approximation needs further investigation. We
argue that the validity is due to the small binding energy for the
deuteron.  From inspection of the next iterations we find additive
corrections only.

Therefore these  assumptions can be loosen up and additional corrections
can be taken into account. This way we have a possibility to calculate
corrections in a more consistent way. This problem is present in both,
electromagnetic and strong processes (for $pd \to pd$ and $pd \to pX$
reactions, for instance).

Although the calculation of these corrections is important to arrive
at definite conclusions about the covariant approach as a whole, our
main issue here is to investigate the correspondence between the
covariant approach and the mesonic exchange currents of the
nonrelativistic approaches. We have shown the analytic correspondence
of these different approaches.

The pair correction plays an essential role in
explaining the experimental data.  An interesting result in this
context  is connected to the form factor entering in the pair current
term. The overall form factor that is consistent with gauge
independence and the relativistic (covariant) expansion neither
coincides with $G_E$ nor with $F_1$ but requires an ``average''
between them.

\section{Acknowledgments}

The authors wish to thank Kaptari L.P., Titov A.I. and especially
Karmanov V.A. for useful and stimulating discussions. We thank Semikh
S.S. and Sus'kov S.Eh. for assistance in the numerical calculations.
The authors are grateful to each others home institutions for extended
hospitality during the respective visits. Part of the work has been
done during the Research Workshop on ``Progress in Current Few Body
Problems'' hosted by the Bogoliubov Laboratory of Theoretical Physics
and supported by the Heisenberg-Landau program.  This work is
partially supported by a grant of the Deutsche Forschungsgemeinschaft
and a RFFI grant of support of leading schools in Russia.


\newpage
\begin{figure}[bp]
  \begin{center}
    \leavevmode
    \psfig{figure=deutBW.eps,width=\textwidth}
 \caption{\label{fig:compare}
Deuteron wave function for the scalar model; exact results
   (left two figures) with $1^{st}$-iteration (right two figures) for
   $^3S_1^{++}$ (upper two figures) and $^3P_1^o$ (lower two figures).
   }
  \end{center}
\end{figure}

\begin{references}
\bibitem{cam82} A. Cambi, B. Mosconi, and P. Ricci, Phys. Rev. Lett
  {\bf 48},  462 (1982).
\bibitem{sch92} M. van der Schaar {\em et al.}, Phys. Rev. Lett. {\bf 68},
  776 (1992), Phys. Rev. Lett. {\bf 66}, 2855 (1991).
\bibitem {lykasov} G.I. Lykasov, Phys. Part. Nucl. {\bf 24}, 59  (1993)
  (Fiz. Elem. Chastits At. Yadra {\bf 24}, 140 (1993)).
\bibitem {kaptari} L.P. Kaptari, A.Yu. Umnikov,  F.C. Khanna, and
  B. K\"ampfer,   Phys. Lett. B {\bf 351}, 400 (1995).
\bibitem{baldin} G.I. Lykasov, Proceedings of the XIII ISHEPP ``Relativistic
  Nuclear Physics and Quantum Chromodynamics'', Dubna, 1996.
\bibitem{burov} V.V.Burov, V.N. Dostovalov and S.{\'E}. Sus'kov,
  Sov. J. Nucl. Phys. {\bf 23}, 317 (1992).
\bibitem{shebeko} A.V. Shebeko, V. Kotlyar, and Yu. Mel'nik, 
  Phys. Part. Nucl. {\bf 26}, 79 (1995).
\bibitem{friar80} J.L. Friar, Phys. Rev. C {\bf 22}, 796 (1980).
\bibitem{adam95} J. Adam Jr., Proc. 14th Int. Conf. on Few Body
  Problems, Williamsburg 1994, edited by F. Gross, AIP Conf. Proc {\bf 334},
 192  (1995). 
\bibitem{ritz96} F. Ritz, H. G\"oller, T. Wilbois, H. Arenh\"ovel,
  Phys. Rev. C {\bf 55}, 2214 (1997). 
\bibitem{kei91} B.D. Keister and W.N. Polyzou, Adv. Nucl. Phys. {\bf 20},
    225 (1991).
\bibitem{dev93} N.K. Devine and S.J. Wallace, Phys. Rev. C {\bf 48},
  R973 (1993).
\bibitem{tjon95} J. Tjon,  Proceedings of the 14th Int. Conf. on Few Body
  Problems, Williamsburg 1994, edited by F. Gross, AIP Conf. Proc {\bf 334},
  177 (1995).
\bibitem{lurie}  D.\,Lurie, {\em Particles and Fields} (Interscience
Publishers, New York, 1968).
\bibitem{IS} C. Itzykson, J.-B. Zuber, {\em Quantum Field Theory}
  (McGraw-Hill, Singapore 1985).
\bibitem{nakanishi} N. Nakanishi, Prog. Theor. Phys. (Kyoto) Suppl. 
 {\bf 43}, 1 (1969).
\bibitem{umn94} A. Yu. Umnikov, L.P. Kaptari,  K.Yu. Kazakov, and
   F.C. Khanna,   Phys. Lett. B {\bf 334}, 163 (1994).
\bibitem{nie96} T. Nieuwenhuis, J.A.Tjon, Few-Body Syst. {\bf 21} 167 (1996).
\bibitem{umn96} Umnikov A.Yu., Z. Phys. A {\bf 357}, 333 (1997).
\bibitem{are94} H. Arenh\"ovel,  Lecture Notes in Physics, Vol. 426,
  eds L. Mathelitsch and W. Plessas (1995) p 1.
\bibitem{edexp} S. Auffret {\it et al.}, Phys. Rev. Lett. {\bf 55}
  1362 (1985); R.G. Arnold {\it et al.}, Phys. Rev. C {\bf 42}, R1 (1990).
\bibitem{edtheor} J.A. Lock and L.L. Foldy, Ann. of Phys. {\bf 93}, 276 (1975).
\bibitem{mathiot} J.-F. Mathiot, Nucl. Phys. {\bf A412}, 201 (1984).
\bibitem{f1ge} S.K. Singh, W. Leidemann, H. Arenh\"ovel, Z. Phys. A {\bf
    331}, 509 (1988).
\bibitem{arenh}    W. Leidemann, and H. Arenh\"ovel, Z. Phys. A {\bf
    326}, 333 (1987).
\bibitem{sch91} R. Schiavilla, and D.O. Riska,
Phys. Rev. C {\bf 43}, 437 (1991).
\bibitem{karmanov} B. Desplanques, V.A. Karmanov, and J.-F. Mathiot,
  Nucl. Phys. {\bf A589}, 697 (1995); J. Carbonell,  B.Desplanques,
  V.A. Karmanov, and J.-F. Mathiot, Phys. Rep. to be published.
\bibitem{kkb} Kaptari L.P., Umnikov A.Yu., S.G. Bondarenko,
  K.Yu. Kazakov, F.C. Khanna, and K\"ampfer B., Phys. Rev. C {\bf 54},
  986 (1996).
\bibitem{fewbody1997} J. Adam Jr., A. Stadler, M.T. Pena, and F. Gross,
  Nucl. Phys. A {\bf 631} 570c (1998); S. Wallace {\em ibid.} 137c.
\bibitem{bbbd:elec} S.G. Bondarenko, V.V. Burov, M. Beyer, and S.M. Dorkin,
MPG-VT-UR 87/96, {\em e-Print Archive: nucl-th/9612047}.
\bibitem{kubis} J.J. Kubis, Phys. Rev. {\bf 6}, 547 (1972).
\bibitem{ZT} M.J. Zuilhof and J.A. Tjon, Phys. Rev. C {\bf 22}, 2369 (1980).
\bibitem{suskov} V.V. Burov, A.A. Goy, S.Eh. Sus'kov, and Yu.V. Chubov,
Sov. J. Nucl. Phys. {\bf 59}, 989 (1996).
\bibitem{NNN6} J.F. Mathiot, Nucl.Phys. {\bf A412}, 201 (1984).
\bibitem{hoehler} G. H\"ohler {\it et al.}, Nucl. Phys. {\bf B114}, 29 (1976).
\bibitem{machl87} R. Machleidt, K. Holinde, and Ch. Elster, Phys.Rep.
  {\bf 149}, 1 (1987).
\bibitem{gari83} M. Gari and U. Kaulfuss, Nucl. Phys. {\bf A408}, 507 (1983).
\bibitem{pair} V.V. Burov, V.N. Dostovalov, and S.Eh.Sus'kov, Czech. J. of
  Phys. {\bf 41}, 1139 (1991).
\bibitem{paris-wf} M. Lacombe {\em et al.}, Phys.Rev. {\bf C21}, 861 (1980).
\end{references}
\end{document}